\documentclass[aps,preprint,showpacs,floatfix]{revtex4}
\usepackage{bm,epsfig}
\begin{document}

\count255=\time\divide\count255 by 60 \xdef\hourmin{\number\count255}
  \multiply\count255 by-60\advance\count255 by\time
 \xdef\hourmin{\hourmin:\ifnum\count255<10 0\fi\the\count255}

\newcommand{\xbf}[1]{\mbox{\boldmath $ #1 $}}

\newcommand{\sixj}[6]{\mbox{$\left\{ \begin{array}{ccc} {#1} & {#2} &
{#3} \\ {#4} & {#5} & {#6} \end{array} \right\}$}}

\newcommand{\ninej}[9]{\mbox{$\left\{ \begin{array}{ccc} {#1} & {#2} &
{#3} \\ {#4} & {#5} & {#6} \\ {#7} & {#8} & {#9} \end{array}
\right\}$}}

\newcommand{\threej}[6]{\mbox{$\left( \begin{array}{ccc} {#1} & {#2} &
{#3} \\ {#4} & {#5} & {#6} \end{array} \right)$}}

\newcommand{\clebsch}[6]{\mbox{$\left( \begin{array}{cc|c} {#1} & {#2} &
{#3} \\ {#4} & {#5} & {#6} \end{array} \right)$}}

\newcommand{\iso}[6]{\mbox{$\left( \begin{array}{cc||c} {#1} & {#2} &
{#3} \\ {#4} & {#5} & {#6} \end{array} \right)$}}

\title{An Identity on SU(2) Invariants}

\author{Herry J. Kwee}
\email{Herry.Kwee@asu.edu}

\author{Richard F. Lebed}
\email{Richard.Lebed@asu.edu}

\affiliation{Department of Physics, Arizona State University, Tempe, AZ
85287-1504}

\date{June 2007}

\begin{abstract}
We prove an identity [Eq.~(1) below] among SU(2) $6j$ and $9j$ symbols
that generalizes the Biedenharn-Elliott sum rule.  We prove the result
using diagrammatic techniques (briefly reviewed here), and then
provide an algebraic proof.  This identity is useful for studying
meson-baryon scattering in which an extra isoscalar meson is produced.
\end{abstract}


\maketitle

\section{Introduction} \label{intro}

In this paper we prove the following identity for a particular sum
over two $9j$ and one $6j$ symbol which, to the best of our knowledge,
seems not to appear previously as a distinct entity in the
literature:
\begin{eqnarray}
\lefteqn{
\sixj{j_1}{j_2}{j_3}{j_4}{j_5}{j_6}
\sixj{j_3}{j_4}{j_5}{\Delta}{j^\prime_5}{j^\prime_4}
\sixj{j^\prime_1}{j^\prime_2}{j_3}{j^\prime_4}{j^\prime_5}{j^\prime_6}
}
&& \nonumber \\
&=& \sum_{k,\ell} [k][\ell] (-1)^\Phi 
\ninej{j_1}{j^\prime_1}{k}{j_6}{j^\prime_6}{\ell}{j_5}{j^\prime_5}
{\Delta}
\ninej{j_2}{j^\prime_2}{k}{j_6}{j^\prime_6}{\ell}{j_4}{j^\prime_4}
{\Delta}
\sixj{j_1}{j_2}{j_3}{j^\prime_2}{j^\prime_1}{k}
\, , \label{newident}
\end{eqnarray}
where $[j] \equiv 2j + 1$ is the multiplicity of the spin-$j$
irreducible representation, and the argument of the phase $(-1)^\Phi$
is given by
\begin{equation} \label{phasearg}
\Phi = j_1 + j_2 + j_6 - j^\prime_4 - j^\prime_5 - j^\prime_6 + k -
\Delta \ .
\end{equation}

This identity is a generalization of the well-known Biedenharn-Elliott
(BE) sum rule [Eq.~(\ref{BE}) below], to which (as we shall show) it
reduces when $\Delta$ is set to zero.

This identity arises when considering the angular momentum coupling
scheme
\begin{equation} \label{coupling}
\begin{array}{ccc}
{\bf j_3} = {\bf j_1} + {\bf j_2} , \ &
{\bf j_3} = {\bf j^\prime_1} + {\bf j^\prime_2} , \ &
{\bf j^\prime_1} = {\bf j_1} + {\bf k} , \\
{\bf j_6} = {\bf j_2} + {\bf j_4} , \ &
{\bf j^\prime_6} = {\bf j^\prime_2} + {\bf j^\prime_4} , \ &
{\bf j_2} = {\bf j^\prime_2} + {\bf k} , \\
{\bf j_5} = {\bf j_3} + {\bf j_4} , \ &
{\bf j^\prime_5} = {\bf j_3} + {\bf j^\prime_4} , \ & 
{\bf j^\prime_4} = {\bf j_4} + {\bf \Delta} \\
{\bf j_5} = {\bf j_1} + {\bf j_6} , \ &
{\bf j^\prime_5} = {\bf j^\prime_1} + {\bf j^\prime_6} , \ &
{\bf j^\prime_5} = {\bf j_5} + {\bf \Delta} , \\
{\bf j^\prime_6} = {\bf j_6} + {\xbf \ell} , \ &
{\bf \Delta} = {\bf k} + {\xbf \ell} .
\end{array}
\end{equation}
The unprimes and primes suggest (for example) initial- and final-state
quantum numbers whose inequality is enforced by a ``spurion'' ${\bf
\Delta}$.  Other physically useful coupling schemes may be obtained by
globally replacing any of these angular momenta {\bf j} by their
time-reversed forms $\tilde{\bf j} \equiv -{\bf j}$, where $\tilde{\bf
j}$ is the angular momentum operator whose eigenstates are related to
those $( \left| j m \right>)$ of ${\bf j}$ by $(-1)^{j+m} \left| j -m
\right>$.  This manipulation gives a well-defined meaning to the
concept of subtracting angular momentum operators~\cite{Edmonds}.

As an explicit physical example using this coupling scheme, consider
meson-baryon scattering ($\phi B \to \phi^\prime B^\prime$) in which
an additional isoscalar meson $f$ is produced with total angular
momentum $J_f$ with respect to the other final-state particles: $\phi
B \to \phi^\prime B^\prime f$.  The specified observables are isospins
and angular momenta: $I_{\phi (\phi^\prime)}$, $J_{\phi
(\phi^\prime)}$ for the mesons (as usual, $J$ denotes spin and orbital
angular momenta combined), and $I_{B (B^\prime)}$, $S_{B (B^\prime)}$
for the baryons.  The total $s$-channel quantum numbers are ${\bf I_s}
\! \equiv \! {\bf I}_\phi \! + {\bf I}_B \! = \! {\bf I}_{\phi^\prime}
\! + {\bf I}_{B^\prime}$, ${\bf J_s} \! = \! {\bf J}_\phi \! + {\bf
S}_B$, ${\bf J^\prime_s} \!  = \! {\bf J}_{\phi^\prime} \! + {\bf
S}_{B^\prime}$, and ${\bf J^\prime_s} \!  = \! {\bf J_s} \! - \! {\bf
J}_f$.  In a chiral soliton model or the $1/N_c$ expansion of QCD, the
vector sum of isospin and angular momentum for each particle assumes
extra significance: The stable baryons (such as nucleons) are
zero-eigenvalue states of the operators ${\bf I}_B \! + \!  {\bf S}_B$
and ${\bf I}_{B^\prime} \! + \!  {\bf S}_{B^\prime}$, and the
scattering is characterized by the ``grand spins'' ${\bf K} \! \equiv
\! {\bf I_s} \! + \! {\bf J_s}$, ${\bf K}^\prime \! \equiv \!  {\bf
I_s} \! + \! {\bf J^\prime_s}$.  The application of
Eq.~(\ref{newident}) arises when one considers processes such as this
not in the $s$-channel but the $t$-channel~\cite{cross}: Then ${\bf
I}_{\phi^\prime} \! = \! {\bf I}_\phi \! + {\bf I_t}$ and ${\bf
J}_{\phi^\prime} \! = \! {\bf J}_\phi \! + {\bf J_t}$.  The full
identification using the notation of Eq.~(\ref{coupling}) is ${\bf
j_1} \! \to \!  {\bf I}_\phi$, ${\bf j_2} \! \to \! {\bf I}_B$, ${\bf
j_3} \! \to \!  {\bf I_s}$, ${\bf j_4} \! \to \! {\bf J_s}$, ${\bf
j_5} \! \to \! {\bf K}$, ${\bf j_6} \! \to \! {\bf J}_\phi$ (and
analogously for the primed ${\bf j}$'s), and ${\bf k} \! \to \! {\bf
I_t}$, ${\xbf \ell} \! \to \!  {\bf J_t}$, and ${\bf \Delta} \! \to \!
-{\bf J}_f$.  The transcription of Eq.~(\ref{newident}) then reads
\begin{eqnarray}
\lefteqn{
\sixj{I_\phi}{I_B}{I_s}{J_s}{K}{J_\phi}
\sixj{I_s}{J_s}{K}{J_f}{K^\prime}{J_s^\prime}
\sixj{I_{\phi^\prime}}{I_{B^\prime}}{I_s}{J_s^\prime}{K^\prime}
{J_{\phi^\prime}}
}
&& \nonumber \\
&=& \sum_{I_t,J_t} [I_t][J_t] (-1)^\Phi 
\ninej{I_\phi}{I_{\phi^\prime}}{I_t}{J_\phi}{J_{\phi^\prime}}{J_t}{K}
{K^\prime}{J_f}
\ninej{I_B}{I_{B^\prime}}{I_t}{J_\phi}{J_{\phi^\prime}}{J_t}{J_s}
{J_s^\prime}{J_f}
\sixj{I_\phi}{I_B}{I_s}{I_{B^\prime}}{I_{\phi^\prime}}{I_t}
\, , \label{example}
\end{eqnarray}
where now $\Phi \! = \! I_\phi \! + \! I_B \! + \! J_\phi \! - \!
J_s^\prime \! - \! K^\prime \! - \! J_{\phi^\prime} \! + \! I_t \! -
\! J_f$.  The three $6j$ symbols on the left-hand side (expressed
solely in terms of $s$-channel quantities) appear as coefficients in
expressions for partial-wave scattering amplitudes for the process
$\phi B \to \phi^\prime B^\prime f$ written in terms of underlying
``reduced'' amplitudes labeled by $K$ values; for example, the case in
which $f$ is absent (effectively, $J_f \! = \! 0$) has been studied
for quite some time~\cite{MP}.  Expressing amplitudes in terms of
$t$-channel quantities [as is manifest on the right-hand side of
Eq.~(\ref{example})] is of particular interest because such amplitudes
scale as $1/N_c^{|I_t \!  - \!  J_t|}$~\cite{cross}, thereby creating
a hierarchy of dominant and subdominant amplitudes in the $1/N_c$
expansion of QCD.

The most illuminating proof of Eq.~(\ref{newident}) uses diagrammatic
techniques, an approach we summarize in Sec.~\ref{diagrammatic}.  We
present the diagrammatic proof in Sec.~\ref{proof}, and finally an
algebraic proof, using standard SU(2) identities, in
Sec.~\ref{algebraic}.

\section{Diagrammatic Method for Coupling Angular Momenta}

\label{diagrammatic}

\subsection{Notation}

Algebraic techniques for manipulating Clebsch-Gordan coefficients
(CGC) to obtain invariants (quantities independent of magnetic quantum
numbers $m$, such as $6j$ and $9j$ symbols) are certainly
straightforward and appear in all standard treatments of the
topic~\cite{Edmonds}.  However, at a certain point of complexity these
techniques become particularly cumbersome, and the bookkeeping
necessary to impose the required identities for simplifying such
expressions [particularly with regard to the numerous phases $(-1)^n$
that arise] becomes increasingly onerous.  A much cleaner strategy is
to use diagrammatic techniques introduced originally by Jucys
(alternate spellings {\it Yutsis}, {\it Iutsis}), Levinson, and
Vanagas (JLV)~\cite{JLV}.  In this method, each angular momentum $j$
is represented as a line, and each vertex represents a $3j$ symbol (or
CGC).  The quantum number $m$ corresponding to $j$ is summed if and
only if line $j$ is internal to the diagram.

The diagrammatic technique is particularly valuable because of two
features: First, the identities involved in combining large complexes
of angular momenta become topological in nature, and the ability to
identify them reduces to one's cunning in picturing how to connect the
lines.  Second, the phases endemic to CGC are incorporated in the
diagrams very neatly (as too are factors of $[j]$, but we do not need
them here), appearing as either signs at the vertices or arrows on the
lines, in the manner described below.  The JLV technique is laid out
pedagogically in the text by Lindgren \& Morrison (LM)~\cite{LM} or
the review by Wormer and Paldus~\cite{WP}.  Here we list only the
features essential for this paper.

\begin{figure}[ht]
\epsfxsize 1.8 in \epsfbox{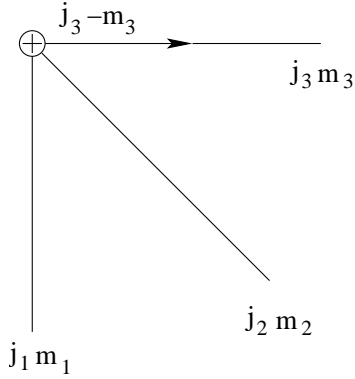}
\caption{Graphical representation of the $3j$ symbol.}
\label{fig:3j}
\end{figure}

For starters, the vertex in Fig.~\ref{fig:3j} represents a $3j$
symbol:
\begin{eqnarray}
\left(\begin{array}{ccc}
j_1 & j_2 & j_3 \\
m_1 & m_2 & -m_3
\end{array}
\right)(-1)^{j_3-m_3},
\end{eqnarray}
with the specific ordering of the $j$'s and $m$'s as
(counter)clockwise defining the vertex orientation as (positive)
negative, as indicated by a sign at the vertex in the diagram.  The
arrow on $j_3$ introduces the phase $(-1)^{j_3-m_3}$; were it pointing
toward the vertex, the phase introduced would be $(-1)^{j_3+m_3}$.
Note that we follow the LM arrow convention, which is opposite that of
JLV convention, since as argued in Ref.~\cite{LM} it is more closely
analogous to the flow of momentum in scattering diagrams.

Several manipulations help simplify such calculations:
\begin{enumerate}
\item Two arrows pointing in opposite direction on the same line can be
removed.
\item Reversing the direction of an arrow introduces an additional
$(-1)^{2j}$.
\item Reversing the orientation of the vertex (changing the sign symbol)
introduces an additional $(-1)^{j_1+j_2+j_3}$.
\item Introducing three arrows all pointing inward or outward at a
vertex does not change the value of the diagram.
\end{enumerate}

\subsection{Invariants}

\begin{figure}[ht]
\epsfxsize 1.8 in \epsfbox{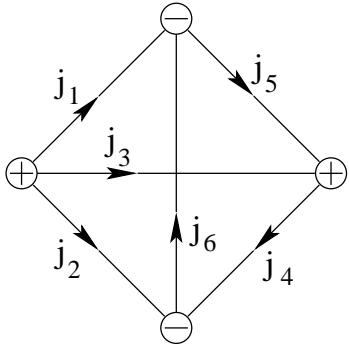}
\caption{Graphical representation of the $6j$ symbol.}
\label{fig:6j}
\end{figure}

\begin{figure}[ht]
\epsfxsize 1.8 in \epsfbox{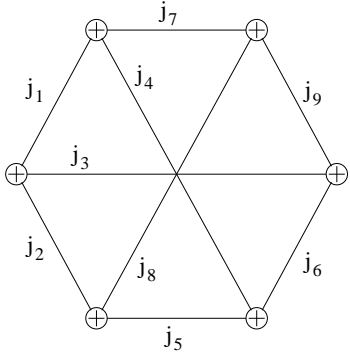}
\caption{Graphical representation of the $9j$ symbol.}
\label{fig:9j}
\end{figure}

The combination of several vertices (with no external lines) forms a
diagram representing a higher-order $3nj$ symbol; for example, the
irreducible combination of 4 vertices, as depicted in
Fig.~\ref{fig:6j}, forms the $6j$ symbol,
\begin{equation}
\sixj{j_1}{j_2}{j_3}{j_4}{j_5}{j_6} ,
\end{equation}
and the irreducible combination of 6 vertices, as depicted in
Fig.~\ref{fig:9j}, forms the $9j$ symbol,
\begin{equation}
\ninej{j_1}{j_2}{j_3}{j_4}{j_5}{j_6}{j_7}{j_8}{j_9} ,
\end{equation}

\subsection{Theorems}

\begin{figure}[ht]
\epsfxsize 5.8 in \epsfbox{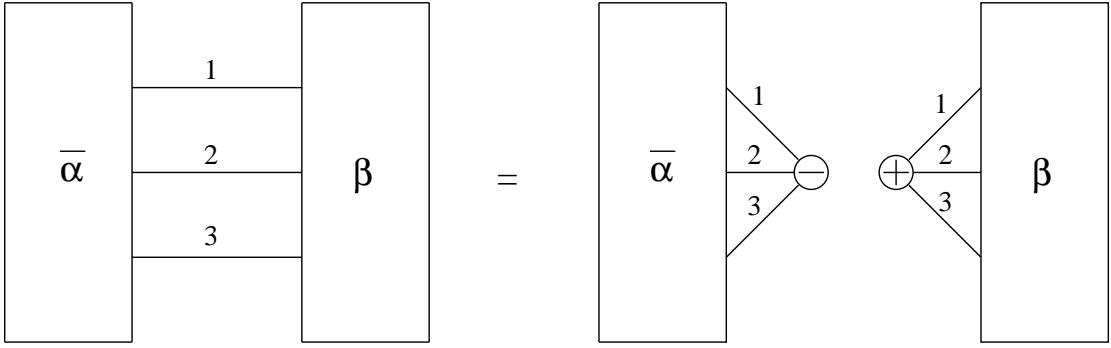}
\caption{The JLV3 theorem.}
\label{fig:JLV3}
\end{figure}

\begin{figure}[ht]
\epsfxsize 4.8 in \epsfbox{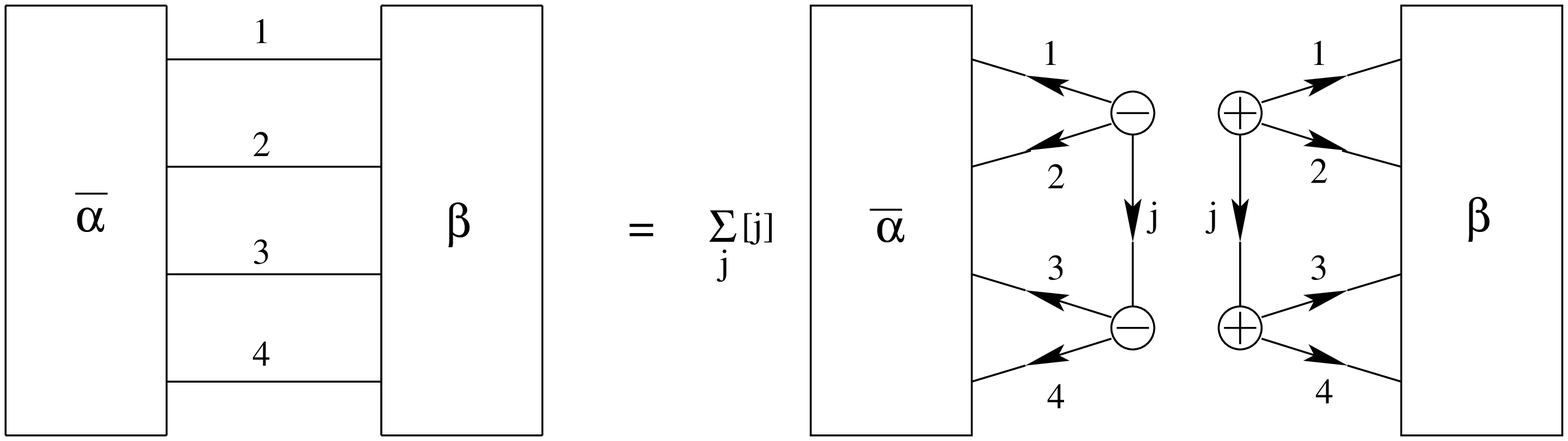}
\caption{The JLV4 theorem.}
\label{fig:JLV4}
\end{figure}

The true power of the JLV approach arises through a serious of
theorems~\cite{JLV,LM}, reminiscent of the factorization theorems of
quantum field theory, that allow one to cut diagrams internally
connected only by a small number $n$ of lines.  Consider a diagram
consisting of two such blocks, $\overline{\alpha}$ and $\beta$, such
that $\overline{\alpha}$ is {\it closed\/} (no external lines) and in
{\it normal form\/} (every internal line on $\overline{\alpha}$
carries an arrow, and any arrows on the lines connecting with $\beta$
are pushed into the block $\beta$).  Then one obtains a series of
theorems JLV$n$, $n \! = \! 1, 2, \ldots$.  Of greatest interest to us
here are JLV3 (depicted in Fig.~\ref{fig:JLV3}) and JLV4
(Fig.~\ref{fig:JLV4}).  JLV3, for example, applied to a system in
which block $\beta$ is empty, turns out to be none other than the
Wigner-Eckart theorem.

\section{Proof of the Identity} \label{proof}

\begin{figure}[ht]
\epsfxsize 5.8 in \epsfbox{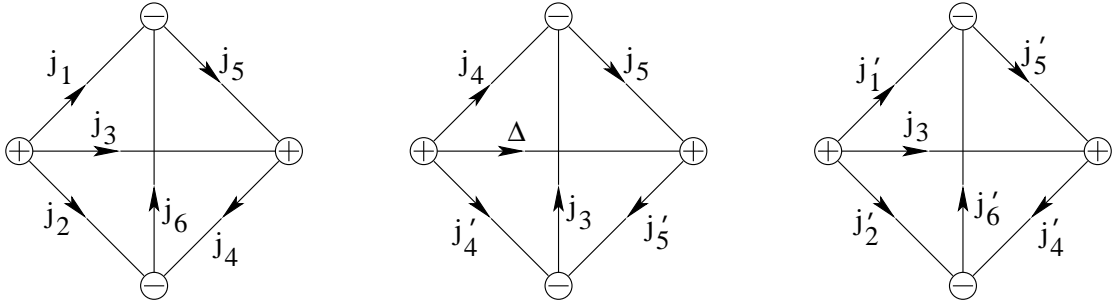}
\caption{The three $6j$ symbols of Eq.~(\ref{three6js}).}
\label{fig:three_6j}
\end{figure}

The Biedenharn-Elliott sum rule~\cite{Edmonds} reads
\begin{eqnarray} \label{BE}
\lefteqn{
\sixj{j_1}{j_2}{j_3}{j_4}{j_5}{j_6}
\sixj{j^\prime_1}{j^\prime_2}{j_3}{j_4}{j_5}{j^\prime_6}
} \nonumber \\ & = &
\sum_{\cal J} (-1)^\sigma [{\cal J}]
\sixj{j_1}{j_6}{j_5}{j^\prime_6}{j^\prime_1}{\cal J}
\sixj{j_2}{j_6}{j_4}{j^\prime_6}{j^\prime_2}{\cal J}
\sixj{j_1}{j_2}{j_3}{j^\prime_2}{j^\prime_1}{\cal J} ,
\end{eqnarray}
where $\sigma$ is the sum of the 9 distinct arguments on the left-hand
side, plus ${\cal J}$.  Here we wish to find the analog of the BE sum
rule for the following three $6j$ symbols, the left-hand side of
Eq.~(\ref{newident}), represented graphically in
Fig.~\ref{fig:three_6j}:
\begin{equation} \label{three6js}
\sixj{j_1}{j_2}{j_3}{j_4}{j_5}{j_6}
\sixj{j_3}{j_4}{j_5}{\Delta}{j^\prime_5}{j^\prime_4}
\sixj{j^\prime_1}{j^\prime_2}{j_3}{j^\prime_4}{j^\prime_5}{j^\prime_6}
\end{equation}

\begin{figure}[ht]
\epsfxsize 4.0 in \epsfbox{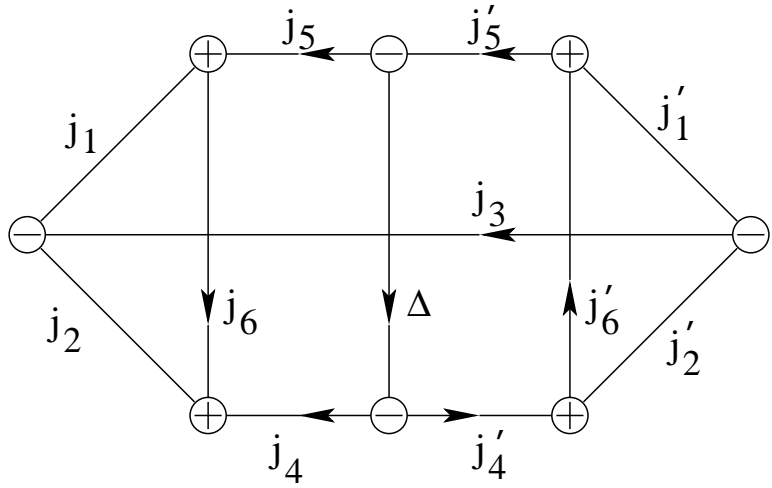}
\caption{Result of combining the three $6j$ symbols of
Eq.~(\ref{three6js}) or Fig.~\ref{fig:three_6j}.}
\label{fig:12j}
\end{figure}

\begin{figure}[ht]
\epsfxsize 4.8 in \epsfbox{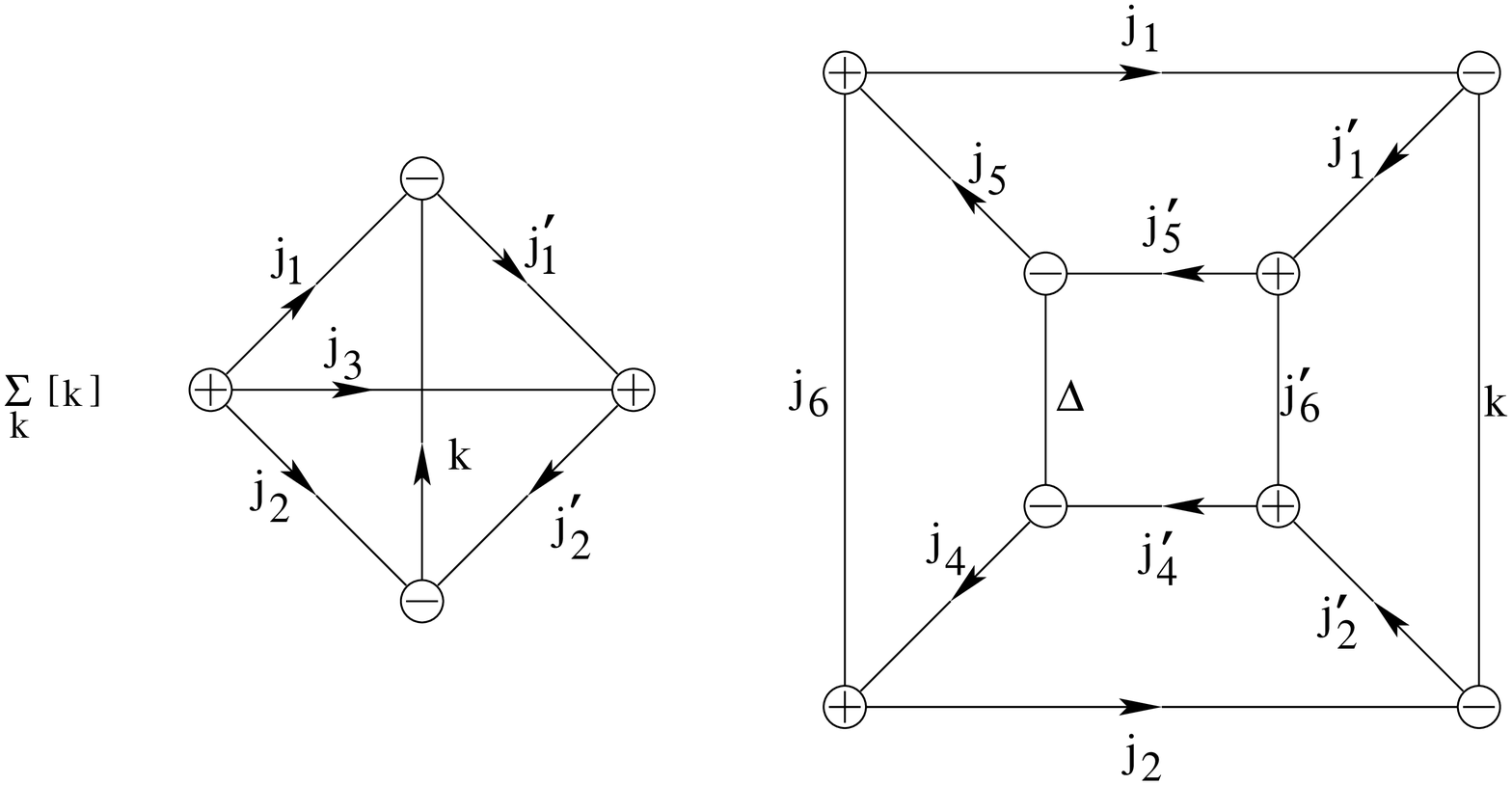}
\caption{The result of applying JLV4 to Fig.~\ref{fig:12j}.}
\label{fig:6j_12j}
\end{figure}

By means of the JLV3 theorem, the three $6j$ symbols of
Eq.~(\ref{three6js}) combine into a diagram with 12 quantum numbers
shown in Fig.~\ref{fig:12j}.  In order to introduce the quantum number
$k$ of Eq.~(\ref{newident}), we cut the diagram on the four lines,
$j_1$, $j^\prime_1$, $j_2$, and $j^\prime_2$ using the JLV4 theorem.
The summed quantum number in this diagram is indeed the desired $k$,
and the result is a $6j$ symbol and a $12j$ {\it symbol of the second
kind\/}, as shown in Fig.~\ref{fig:6j_12j}.  Since the latter object
is surely obscure to most readers, we pause to point out that, while
higher $3nj$ symbols are not difficult to generate and manipulate
using the diagrammatic approach, and while they possess remarkable and
intricate symmetry properties, they may always be reduced to products
over convenient sums of $6j$ and $9j$ symbols using the JLV
theorems~\cite{BC,MB}.  One other fine point is that the attractive
square diagram in Fig.~\ref{fig:6j_12j} actually differs from the true
$12j$ symbol by a phase $(-1)^{j_1 - j_2 + j^\prime_4 - j^\prime_5}$,
but this distinction is purely formal; like Eq.~(\ref{newident}), the
diagram as depicted is symmetric upon exchange of primed and unprimed
quantum numbers.

\begin{figure}[ht]
\epsfxsize 6.45 in \epsfbox{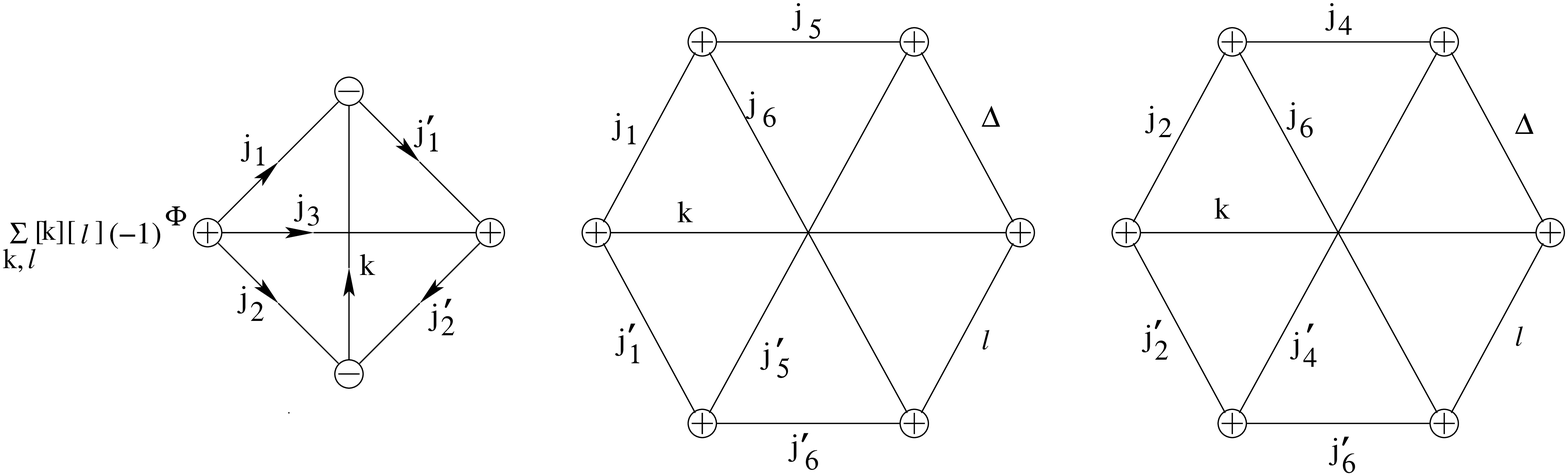}
\caption{The result of applying JLV4 to Fig.~\ref{fig:6j_12j},
which is expressed algebraically in Eq.~(\ref{newident}).}
\label{fig:6j_two_9j}
\end{figure}

To introduce the quantum number $\ell$, we make another four-line cut,
here on $j_6$, $j^\prime_6$, $k$, and $\Delta$ in the $12j$
symbol of Fig.~\ref{fig:6j_12j}.  The result of this action is shown
in Fig.~\ref{fig:6j_two_9j}.  The hexagonal figures are none other
than standard $9j$ symbols in canonical JLV form.

In fact, we have expressed a particular product of three $6j$ symbols
[Eq.~(\ref{three6js})] as a $6j$ and two $9j$ symbols summed over two
new quantum numbers, $k$ and $\ell$.  To repeat Eq.~(\ref{newident}),
\begin{eqnarray}
\lefteqn{
\sixj{j_1}{j_2}{j_3}{j_4}{j_5}{j_6}
\sixj{j_3}{j_4}{j_5}{\Delta}{j^\prime_5}{j^\prime_4}
\sixj{j^\prime_1}{j^\prime_2}{j_3}{j^\prime_4}{j^\prime_5}{j^\prime_6}
}
&& \nonumber \\
&=& \sum_{k,\ell} [k][\ell] (-1)^\Phi 
\ninej{j_1}{j^\prime_1}{k}{j_6}{j^\prime_6}{\ell}{j_5}{j^\prime_5}
{\Delta}
\ninej{j_2}{j^\prime_2}{k}{j_6}{j^\prime_6}{\ell}{j_4}{j^\prime_4}
{\Delta}
\sixj{j_1}{j_2}{j_3}{j^\prime_2}{j^\prime_1}{k}
\, ,
\end{eqnarray}
with the phase given by
\begin{equation}
\Phi = j_1 + j_2 + j_6 - j^\prime_4 - j^\prime_5 - j^\prime_6 + k -
\Delta \ .
\end{equation}
While $\Phi$ is not primed-unprimed symmetric, neither are the two new
$9j$ symbols, because switching their initial and final quantum
numbers requires column exchanges; when the necessary permutations are
taken into account, it is quite straightforward to show this explicit
symmetry.  To the best of our knowledge, Eq.~(\ref{newident}) actually
represents a new SU(2) identity~\cite{why_new}, one that reduces for
$\Delta \! = \! 0$ to the BE sum rule.  This reduction becomes
apparent when one simplifies using special cases
\begin{equation}
\sixj{j_3}{j_4}{j_5}{0}{j^\prime_5}{j^\prime_4} =
\frac{(-1)^{j_3 + j_4 + j_5}}{\sqrt{[j_4][j_5]}}
\delta_{j^{\vphantom\dagger}_4 j^\prime_4}
\delta_{j^{\vphantom\dagger}_5 j^\prime_5} ,
\end{equation}
\begin{equation}
\ninej{j_1}{j^\prime_1}{k}{j_6}{j^\prime_6}{\ell}{j_5}{j^\prime_5}{0}
= \frac{(-1)^{j^\prime_1 + j_6 + j_5 + k}}{\sqrt{[k][j_5]}}
\delta_{j^{\vphantom\dagger}_5 j^\prime_5}
\delta_{k^{\vphantom\dagger}_{\vphantom\bullet}
\ell^{\vphantom\dagger}_{\vphantom\bullet}}
\sixj{j_1}{j_6}{j_5}{j^\prime_6}{j^\prime_1}{k} ,
\end{equation}
and
\begin{equation}
\ninej{j_2}{j^\prime_2}{k}{j_6}{j^\prime_6}{\ell}{j_4}{j^\prime_4}{0}
= \frac{(-1)^{j^\prime_2 + j_6 + j_4 + k}}{\sqrt{[k][j_4]}}
\delta_{j^{\vphantom\dagger}_4 j^\prime_4}
\delta_{k^{\vphantom\dagger}_{\vphantom\bullet}
\ell^{\vphantom\dagger}_{\vphantom\bullet}}
\sixj{j_2}{j_6}{j_4}{j^\prime_6}{j^\prime_2}{k} ,
\end{equation}
in which case both $k$ and $\ell$ reduce to ${\cal J}$ of
Eq.~(\ref{BE}).

\section{Algebraic Proof} \label{algebraic}

Equation~(\ref{newident}) is also fairly straightforward to verify
algebraically, once the right-hand side is known.  Here we reduce
this side of the equation.  We use the symmetry properties of $6j$
symbols and $9j$ symbols: $6j$ symbols are invariant under the
permutation of any two columns or under the exchange of upper and
lower entries for any two columns, while $9j$ symbols are invariant
under even permutations of any two rows or columns.  We also use the
BE sum rule, Eq.~(\ref{BE}).  Furthermore, $9j$ symbols may be
expanded in terms of $6j$ symbols using the standard
identity~\cite{Edmonds}
\begin{equation} \label{9jexpand}
\ninej{j_6}{j^\prime_6}{\ell}{j_4}{j^\prime_4}{\Delta}{j_2}
{j^\prime_2}{k}
= \sum_x (-1)^{2x} [x] \sixj{j_6}{j_4}{j_2}{j^\prime_2}{k}{x}
\sixj{j^\prime_6}{j^\prime_4}{j^\prime_2}{j_4}{x}{\Delta}
\sixj{\ell}{\Delta}{k}{x}{j_6}{j^\prime_6} ,
\end{equation}
where the $9j$ symbol is equivalent to the second one in
Eq.~(\ref{newident}).  The first $9j$ symbol of Eq.~(\ref{newident})
and the last $6j$ symbol of Eq.~(\ref{9jexpand}) (arguments
rearranged), along with their sum over $\ell$, may be re-expressed
using another standard identity~\cite{Edmonds}:
\begin{equation} \label{Edmonds6.4.8}
\sum_{\ell} [\ell]
\ninej{j_6}{j^\prime_6}{\ell}{j_5}{j^\prime_5}{\Delta}{j_1}
{j^\prime_1}{k} \sixj{j_6}{j^\prime_6}{\ell}{\Delta}{k}{x} = (-1)^{2x}
\sixj{j_5}{j^\prime_5}{\Delta}{j^\prime_6}{x}{j^\prime_1}
\sixj{j_1}{j^\prime_1}{k}{x}{j_6}{j_5} .
\end{equation}
The phase $(-1)^{2x}$ cancels between Eqs.~(\ref{9jexpand}) and
(\ref{Edmonds6.4.8}), and the remaining expression reads
\begin{equation} \label{preBE}
\sum_x [x] \sum_k [k] (-1)^\Phi 
\sixj{j^\prime_6}{j^\prime_4}{j^\prime_2}{j_4}{x}{\Delta}
\sixj{j_5}{j^\prime_5}{\Delta}{j^\prime_6}{x}{j^\prime_1}
\sixj{j_6}{j_4}{j_2}{j^\prime_2}{k}{x}
\sixj{j_1}{j^\prime_1}{k}{x}{j_6}{j_5}
\sixj{j_1}{j_2}{j_3}{j^\prime_2}{j^\prime_1}{k} .
\end{equation}

The summed angular momentum $k$ appears only in the last three of
these $6j$ symbols, and the phase argument $\Phi$
[Eq.~(\ref{phasearg})] conveniently contains a factor of $k$,
suggesting that they can be simplified using the BE sum rule.  Indeed,
writing the sum of the ten angular momenta in the last three $6j$
symbols as $\sigma$, one has
\begin{equation} \label{firstBE}
\sum_k (-1)^\sigma [k]
\sixj{j_2}{j_6}{j_4}{x}{j^\prime_2}{k}
\sixj{j_6}{j_5}{j_1}{j^\prime_1}{k}{x}
\sixj{j_2}{j_1}{j_3}{j^\prime_1}{j^\prime_2}{k} =
\sixj{j_1}{j_2}{j_3}{j_4}{j_5}{j_6}
\sixj{j_4}{j_5}{j_3}{j^\prime_1}{j^\prime_2}{x} .
\end{equation}
The first $6j$ symbol appears on the left-hand side of
Eq.~(\ref{newident}).  The remaining factors to be simplified (those
containing $x$) are the first two $6j$ symbols in Eq.~(\ref{preBE}),
the second in Eq.~(\ref{firstBE}), and the phase $(-1)^{\Phi-\sigma}
\! = \!  (-1)^{\sigma - \Phi}$ (since $\Phi$ and $\sigma$ are integers)
$\equiv \! (-1)^{\tilde\sigma}$.  The BE sum rule again simplifies the
expression:
\begin{equation} \label{secondBE}
\sum_x (-1)^{\tilde\sigma} [x]
\sixj{j_4}{\Delta}{j^\prime_4}{j^\prime_6}{j^\prime_2}{x}
\sixj{\Delta}{j^\prime_5}{j_5}{j^\prime_1}{x}{j^\prime_6}
\sixj{j_4}{j_5}{j_3}{j^\prime_1}{j^\prime_2}{x} =
\sixj{j_3}{j_4}{j_5}{\Delta}{j^\prime_5}{j^\prime_4}
\sixj{j^\prime_1}{j^\prime_2}{j_3}{j^\prime_4}{j^\prime_5}{j^\prime_6}
,
\end{equation} 
since $\tilde\sigma$ is easily shown to be the sum of the ten angular
momenta on the left-hand side.  These two $6j$ symbols, along with the
first from Eq.~(\ref{firstBE}), complete the left-hand side of
Eq.~(\ref{newident}), and hence complete the proof.

\section*{Acknowledgments}
This work was supported by the NSF under Grant No.\ PHY-0456520.


\begin{thebibliography}{99}
\bibitem{why_new}
Similar identities appear in the literature, but Eq.~(\ref{newident})
appears to be distinct.
\end{thebibliography}
\end{document}